\documentclass[journal]{IEEEtran}

   \ifCLASSINFOpdf   
\else
\fi
\usepackage{times,amsmath,epsfig,latexsym,amssymb,psfrag}
\usepackage{booktabs}
\usepackage{caption}
\usepackage{subcaption}
\usepackage{color}
\usepackage{setspace}
\usepackage{mathrsfs}
\usepackage{enumitem}
\usepackage[ruled,vlined]{algorithm2e}
\usepackage{graphicx}
\usepackage{epstopdf} 
\usepackage{amsmath}
\usepackage{mathtools}
 \usepackage{cite}
\DeclareMathAlphabet{\mathpzc}{OT1}{pzc}{m}{it}
 \newtheorem{theorem}{Theorem}[section]
 
 \newtheorem{proposition}[theorem]{Proposition}

 \newcommand{\qed}{\nobreak \ifvmode \relax \else
 	\ifdim\lastskip<1.5em \hskip-\lastskip
 	\hskip1.5em plus0em minus0.5em \fi \nobreak
 	\vrule height0.75em width0.5em depth0.25em\fi}
 
\begin{document}
 \title{Grant-free Radio Access IoT Networks: \\ Scalability Analysis in Coexistence Scenarios}
\author{\IEEEauthorblockN{Meysam Masoudi\IEEEauthorrefmark{2}, Amin Azari\IEEEauthorrefmark{2}, Emre Altug Yavuz\IEEEauthorrefmark{3}, and Cicek~Cavdar\IEEEauthorrefmark{2}}
\\	\IEEEauthorblockA{\IEEEauthorrefmark{2} Communication Systems Department, KTH Royal Institue of Techonology, Sweden}
	\\ \IEEEauthorblockA{\IEEEauthorrefmark{3} Ericsson AB, Stockholm, Sweden}
\\Email:\{masoudi,aazari,cavdar\}@kth.se, emre.yavuz@ericsson.com
	\thanks{This study is supported by EU Celtic Plus Project SooGREEN Service Oriented Optimization of Green Mobile Networks.}}

\maketitle


\begin{abstract}
IoT networks with grant-free radio access,  like SigFox and LoRa, offer low-cost durable  communications over  unlicensed band. These networks  are becoming more and more popular due to the ever-increasing need for ultra durable, in terms of battery lifetime, IoT networks.  Most studies evaluate the system performance assuming single radio access technology deployment. In this paper, we study the impact of coexisting competing radio access technologies on the system performance. Considering $\mathpzc K$ technologies, defined by time and frequency activity factors, bandwidth, and power, which share a set of radio resources, we derive closed-form expressions for the successful transmission probability, expected battery lifetime, and experienced delay as a function of distance to the serving access point. Our analytical model, which is validated by  simulation results,
provides a tool to evaluate  the coexistence
scenarios and analyze how introduction of a new coexisting
technology may degrade the system performance in terms of
success probability and battery lifetime. We further investigate solutions in which this destructive effect could be compensated, e.g., by densifying the network to a certain extent  and utilizing joint reception. 
\end{abstract}
\IEEEpeerreviewmaketitle

\section{Introduction} 
The number of physical objects  being connected to the Internet is growing at an unprecedented rate, which is realizing the idea of the Internet of Things (IoT) or the Internet of everything. As one of the  major drivers of 5G, it is important to provide (i) scalable (ii) low-cost and (iii) ultra-durable connectivity for the future IoT networks
 \cite{5g_iot}.  Supporting the connectivity over future cellular networks has been a main study item in 3GPP, and several  revolutionary and evolutionary connectivity solutions have been proposed/standardized \cite{5g_iot}. However, the existing solutions do not consider jointly the scalability, durability, and  cost. 

The IoT devices are supposed to send short payload-size packets sporadically \cite{miao2016MAC}. As the number of connected devices increases excessively, the signal and control overhead becomes a burden for the conventional grant-based radio access protocol commonly used in cellular networks due to the  excessive control signaling \cite{du2017efficient}. As a promising solution, grant-free radio access  has attracted a great deal of attention in recent years. In grant-free access, once a packet is triggered at the device, it is transmitted without any handshaking or authentication process. Several existing IoT technologies benefit from a grant-free radio access for providing low-cost long-battery lifetime connectivity, including SigFox and LoRa \cite{lif_com}. The common characteristic of these IoT technologies that distinguishes them from existing WiFi and cellular solutions, used for short-range high-bandwidth connectivity and   long-range mobile connectivity respectively, are leveraging narrowband communication to cover a large area with minimum possible power consumption at the devices. The low-power wide-area (LPWA) networks  are expected to share 60 percent of the IoT  market among themselves, a number that is expected to grow over time, and hence {the competition  between LPWA technologies is becoming intense \cite{all_comp}}. Regarding the increasing number of IoT technologies aiming at providing large-scale IoT connectivity by reusing a set of radio resources, it is of paramount importance to investigate the mutual impacts of coexisting technologies on each other.

\subsection{Literature Study}
 In \cite{all_comp} and \cite{bid}, major IoT solutions over licensed and unlicensed bands have been introduced, and their challenges in providing massive IoT connectivity have been figured out. 
 A thorough battery lifetime analysis   for unlicensed band solutions, including IEEE 802.15.4, BLE, SigFox and LoRa, has been presented in \cite{life_all} based on the physical layer characteristics, i.e. operation protocols, connectivity states,  and consumed energy in each state. Among solutions over unlicensed spectrum, SigFox and LoRa, as introduced in \cite{all_comp}, are dominant solutions.  Performance limits of LoRa have been investigated in \cite{lora_sca,lora_exp,lora_int,lora_lim}. In \cite{lora_sca}, scalability of single-gateway LoRa  network has been investigated, and it has been shown that as the number of end-nodes 
 increases, the impact on co-spreading factor increases, and hence, network becomes interference limited. Experimental results on the impact of interference from other LoRa nodes have been presented in \cite{lora_int}.  In \cite{lora_lim}, performance limits of LoRa have been discussed, and it has been shown that besides the aforementioned limits, regulations govern the ISM band, e.g., duty cycle of operation, also limit the scalability of LoRa networks. The authors of \cite{int1,int2}  experimentally  evaluated the impact of potential  interfering technologies reside in the ISM band.  Their results illustrate a significant impact of interference from IoT devices already installed in smart homes, business parks, and etc., on the performance of LoRa and SigFox communications.  In \cite{gf}, the authors aim at bridging among solutions in the licensed and unlicensed band by presenting a grant-free access scheme over licensed spectrum for long battery lifetime demanding devices.



The literature study reveals that most previous studies have been focusing on the investigation of performance limits for single technology scenarios, and hence, the impact of coexistent technologies with partial time/frequency overlapping has been neglected. While the literature study of interference management for cellular and WiFi networks is mature \cite{khodet}, there is a crucial need to investigate the interference impact for large-scale, heterogeneous, and short-packet communication, which is the focus of this work. The main contributions of this paper are as follows. 
\begin{itemize} 
\item Deriving the closed-form expressions for the  probability of successful transmission, battery lifetime, and delay for a network in which $\mathpzc K$ heterogeneous technologies, defined by their transmit powers, time, and frequency activity factors, are reusing radio resources. Investigating the performance impact of coexisting technologies. 
\item Realizing joint reception as a solution to compensate the degradation due to the interference from other interfering technologies. Analyzing performance  enhancement of the network. Figuring out the limits on the performance improvement.
\end{itemize}

The remainder of this paper is organized as follows. 
In Section \ref{sectionII}, the system model is presented. In Section \ref{sectionIII}, closed-form expressions for the key performance indicators (KPIs) are derived. The performance evaluation is presented in Section \ref{sectionIV}. Concluding remarks are summarized in Section \ref{sectionV}.
 
 \section{System Model}\label{sectionII}
 Assume a network with a massive number of IoT devices from $\mathpzc K$ different technologies which are randomly distributed according to a spatial Poisson point process 
 (PPP). Inside each technology, the pattern of packet generation and  shared spectrum usage across different devices may differ from one another. Then, we define $K$ classes of devices in the network, where $K\ge \mathpzc K$. By a class, we mean devices with a common pattern of  shared spectrum usage, where the pattern includes time-frequency pattern of transmitted packets, range of carrier frequency, transmit power, and rate of packet generation at devices. To collect data from these devices, access points (APs) are deployed  in the interest area. The density of  APs and devices for $i$th class is denoted by $\lambda_{i, a}$ and $\lambda_{i,b }$ respectively. Denote the carrier frequency and  time-frequency support of each packet from a class $i$ device,  by $f_i$ and $\omega_i\times T_i$ Hz$\times$sec,  in which $f_{i,\text{min}}$$\le$$f_i$$\le$$f_{i,\text{max}}$\footnote{The region for carrier frequency captures two facts. First, most IoT devices have cheap oscillators, and hence, the carrier frequency of the transmitted signals drifts from lower and higher frequency ($f_{i,\text{min}}$,$f_{i,\text{max}}$) \cite{gf}. Second, in some IoT technologies like SigFox, IoT devices randomly change the carrier frequency in consecutive transmissions to make the communication robust.}. Also, $\mathcal T_i$ denotes  the average time between two consecutive packet transmissions of a class $i$ device. As in \cite{2d,gf}, we assume that the transmitted energy  is uniformly distributed over its time-frequency support, i.e., over a rectangle of size $\omega_i\times T_i$ Hz$\times$sec for class $i$. The channel gain consists of pathloss with the pathloss exponent of $\alpha$, and Rayleigh fading. The required signal to interference and noise ratio (SINR) threshold for successful signal decoding at an AP is denoted by $\gamma_{\text{th}}$.
\subsubsection*{Problem Description and KPIs} 
The main goal of this paper is modeling the received interference from co-existing technologies, evaluating  the system performance in presence of such interference, and finding solutions to compensate for the performance degradation due to the interference. The main KPIs of interest in this work are as follows.
   
 \subsubsection{Battery lifetime}
  Battery lifetime measures the time span between deployment of a device and when the device has its battery drained. Regarding the fact that most of IoT devices are battery driven,  long battery lifetime is of great importance in most  IoT applications. If batteries of IoT devices need to be replaced frequently, the human intervention, and hence maintenance cost,  will be high, which limits the scalability of IoT networks.
 
 \subsubsection{Experienced delay}
 The experienced delay is defined as a delay from having packet ready at the device to successful reception of data at the access point. Characterizing the statistics of the experienced delay is important for IoT applications with stringent delay requirement.
 
 
   \begin{table}[t!]
	\centering \caption{Frequently used symbols.}\label{AnaR}
	\begin{tabular}{p{2.7 cm}p{5.32 cm}}\\
		\toprule[0.5mm]
		{\it Symbol}&{\it Definition}\\
		\midrule[0.5mm]
		 $i$ and $j$  &  The class indexes\\
$G_{X}(x)$$\buildrel \Delta \over=$$pr(X$$\le$$ x)  $& CDF of random variable $X$ \\
$ \mathcal G_X(x)$&PDF of random variable $X$\\
$f_i\in \{f_{i,\text{mn}},f_{i,\text{mx}}\}$ & Carrier frequency \\
$K$& Number of classes of devices\\
$\lambda_{i,a}, \lambda_{i,b}$ & Density of APs and devices\\
$\xi_{i,j}$& Time activity factor\\
$\upsilon_{i,j}$& Frequency activity factor\\
$T_i$& Transmission time for a packet\\
$1/\mathcal T_i$& Generation rate of packets\\
$\gamma$, $\gamma_{\text{th}}$& SINR, minimum required SINR\\
$\mathcal P_{\text{sc}}$& Probability of successful transmission\\
$w_i$&Signal bandwidth\\
$\alpha, \sigma$& Pathloss exponent, $2/\alpha$\\
$\mathcal L_{\mathcal I_j}(s)$& Laplace functional of interference ($\mathcal I_j$)\\
$\mathbb E(\cdot)$& Expectation operator\\
$E_j$ & Battery capacity in Joules\\
$ \mathcal E_{j,a}, \mathcal E_{j,b}$& Energy consumption of device in 1 reporting period and AP in unit time\\ 
$p^{\text{av}}_{m,j}$& Probability of availability of $m$th AP\\
$P_j$& Transmit power\\ 
		\bottomrule[0.5mm]
	\end{tabular}
\end{table}

 \section{Theoretical Analysis}\label{sectionIII}
 \subsection{Analytical Modeling of Transmission Success Probability}
   \begin{figure}[t!]
 	\includegraphics[scale=.5]{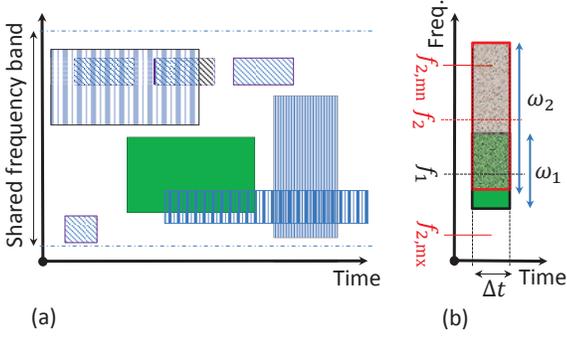}
 	\caption{Grant-free communications: (a) partial time-frequency overlapping of packets, and (b) average received power from an interfering device.}
 	\label{fint}
 \end{figure}
In order to derive the success probability,  we first analyze the case in which devices and APs have been paired, i.e., the transmitted packets from each device are  received just by one AP. Then in Section \ref{mul}, we extend our analysis to the multi-AP scenario, in which several received copies of a packet in different APs are combined for performance enhancement.
 Let us first assume that each class $j$ device has been paired with only one class $j$ AP which is always available to receive data. The duty cycle of class $i$ devices relative to class $j$ devices is denoted by $\mathcal Z(i,j) T_i/\mathcal T_i$. Where $\mathcal Z(i,j)=1$ when  class $i$ and class $j$ refer to two different technologies\footnote{Note: two classes may refer to the same technology, e.g. two SigFox networks with different message generation rates, or two different technologies, e.g. SigFox and LoRa.}, e.g. SigFox and LoRa,  and $\mathcal Z(i,j)$ equals to the inverse of number of  available orthogonal channels and codes in the technology otherwise\footnote{For example in LoRa, we usually have 3 orthogonal channels and 7 spreading factor to select for sending data \cite{all_comp}.}. Then, the transmission success probability, $ \mathcal P_{\text{sc}} $, for a class $j$ device at distance $d$ from the serving AP, is defined as:
 \begin{eqnarray}\label{PSuccess}
 	\mathcal P_{\text{sc}}(j,d,\gamma_{\text{th}}) \triangleq pr (\gamma\geq \gamma_{\text{th}}),
 \end{eqnarray}
 where $ \gamma $ is signal to noise and interference ratio, and is calculated as $\gamma = {\mathcal S}/{[\mathcal I + \mathcal N]}$, in which $ \mathcal S $, $ \mathcal I $, and $ \mathcal N $ account for signal, aggregated interference, and noise powers.  
 We start by the following proposition. 
 
 \begin{proposition} 
 	The success probability  for a packet transmitted by a device from class $j$ operating on carrier frequency $f_j$ at distance $d$ from the AP, is:
 	\begin{align}
 		& 		\mathcal P_{\text{sc}}(j,d,\gamma_{\text{th}},f_j)  {=} \exp( {-} {\gamma_{\text{th}} d^{\alpha}\mathcal N}/{P_j})\times \label{PsuccessCalc0}\\
 		&\hspace{3mm}\prod\nolimits_{i=1}^{K}\exp( {-}\xi_{i,j}\lambda_{i,b} \pi [ {\gamma_{\text{th}} \upsilon_{i,j} P_i}/{P_j}]^\sigma \mathbb E(h^\sigma)\Gamma(1-\sigma) {d^2} ),\nonumber
 	\end{align}
 	in which $\sigma=2/\alpha$, $\xi_{i,j}=\mathcal Z(i,j)T_i/\mathcal T_i $ is the time activity factor of class $i$ devices, {$\mathbb{E}$ stands} for expectation, $P_i$ is the transmit power of each class $i$ device, $\upsilon_{i,j} =\mathcal F(f_{i,\text{mn}},f_{i,\text{mx}},\omega_i,\omega_j,f_j) $ accounts for the expected overlap between packets from $i$ and $j$ classes, and will be defined in Proposition \ref{pr2}.  
 	
 \end{proposition}

 \begin{IEEEproof}
 	Starting from the definition of success probability, we have
 	\begin{align}
 		\mathcal P_{\text{sc}}(j,d,\gamma_{\text{th}},f_j) =&pr(\gamma\geq \gamma_{\text{th}})=
 		pr(\mathcal S> \gamma_{\text{th}}(\mathcal N+\mathcal I_{j}) )\nonumber\\
 		\buildrel (a) \over = &\exp(-{\gamma_{\text{th}} d^{\alpha}\mathcal N}/{P_j}) \mathbb{E}_{\mathcal I_{j}}(e^{-{\gamma_{\text{th}}d^{\alpha}\mathcal I_{j}}/{P_j}})\label{a1}\\
 		=&\exp(- {\gamma_{\text{th}} d^{\alpha}\mathcal N}/{P_j})\mathcal{L}_{\mathcal I_{j}}(s)\arrowvert_{s={\gamma_{\text{th}}d^{\alpha}}/{P_j}},\label{PsuccessCalc}
 	\end{align}
 	where $\mathcal I_{j}$ represents the average aggregated received interference  on  the tagged packet from a class $j$ device and $(a)$ is due to the independence of noise and interference, the first and the second terms in  \eqref{a1}, respectively.  $\mathcal L_{\mathcal I_j}(s)$ is the Laplace functional of interference \cite{Martinbook}. 
 	As $\mathcal I_j$ includes potential interference from all $K$ classes, we have \begin{equation}\label{eq2t}\mathcal I_{j} = \sum\nolimits_{i=1}^{K} \mathcal I_{i,j},\quad \mathcal I_{i,j}=\sum\nolimits_{x \in \Phi_i} h \ell{(x)}\upsilon_{i,j} P_i.\end{equation}
 	In this expression, $\Phi_i $ is the set containing the locations of interfering class $j$ devices, $h$ is fading, 
 	$\upsilon_{i,j}$ represents the length of the overlapped frequency band between packets from class $i$ and class $j$ devices, and $ \ell(x) $ is pathloss component.
 	$ \mathcal L _{\mathcal I_{i,j}}(s)$  is defined as \cite{Martinbook}: 
 	\begin{align}
 		\mathcal{L}_{\mathcal I_{i,j}}(s) &\triangleq \mathbb{E}(\exp({-s \mathcal I_{i,j}}))\nonumber\\
 		&=\mathbb{E}\big(\prod\nolimits_{x\in\Phi_i}\exp(-s\upsilon_{i,j}P_i hr^{-\alpha})\big),
 	\end{align}
 	where the expectation is taken over both $ \Phi_i $ and $ h$. Also, in this expression $\ell(x)=r^{-\alpha}$ is assumed, where $r$ is the distance between interfering class $i$ device located at $x\in\Phi$, and the origin.
 	Since fading is independent of the point process, the expectation operator can be moved inside the product, then we have: 
 	\begin{align}
 		\mathcal{L}_{\mathcal I_{i,j}}(s) &\text{=} \exp\big(\text{-}\int^{\infty}_{0} \mathbb{E}([1\text{-}\exp({{\text{-}s\upsilon_{i,j}P_ihr^{-\alpha}}})])\xi_{i,j} \lambda_{i,b} 2\pi r dr \big),\nonumber\\
 		&\text{=}  \exp(-\xi_{i,j}\lambda_{i,b} \pi \mathbb{E}(h^\sigma) \Gamma(1-\sigma)[s\upsilon_{i,j}P_i]^{\sigma}).	 \label{li}  	  
 	\end{align}
 	Using  \eqref{eq2t} and \eqref{li}, one can derive $\mathcal{L}_{\mathcal I_{j}}(s)$ as:
 	\begin{eqnarray}\label{Laplace_final}
 		\mathcal{L}_{\mathcal I_{j}}(s)  = \prod\nolimits_{i=1}^{K}\exp(-\xi_{i,j}\lambda_{i,b} \pi \mathbb{E}(h^\sigma) \Gamma(1-\sigma)[s\upsilon_{i,j}P_i]^{\sigma}).
 	\end{eqnarray}
 	By combining \eqref{PsuccessCalc} and \eqref{Laplace_final}, the success probability is derived as presented  in  \eqref{PsuccessCalc0}.
 \end{IEEEproof}

 	For Rayleigh distributed fading, the success probability in  \eqref{PsuccessCalc0} is rewritten as:
 	\begin{align}
 		\mathcal P_{\text{sc}}&(j,d,\gamma_{\text{th}},f_j)  {=}\exp( {-} {\gamma_{\text{th}} d^{\alpha}\mathcal N}/{P_j})
 		\nonumber\\
 		&\prod\nolimits_{i=1}^{K}\exp({-}\xi_{i,j}\lambda_{i,b} \pi [ {\gamma_{\text{th}} \upsilon_{i,j} P_i}/{P_j}]^\sigma  {d^2}/{\text{sinc}(\sigma)}).\label{PsuccessCalc0_R}
 	\end{align} 

%
 \begin{proposition} \label{pr2}
 	Assume two devices transmit data simultaneously, as depicted in Fig. \ref{fint}.b, with transmit powers of $P_1$ and $P_2$, and carrier frequencies of $f_1$ and $f_2$, where $f_2$ is a random variable in 
 	$[f_{2,\text{mn}},f_{2,\text{mx}}]$, and follows a general distribution with cumulative distribution function (CDF) of $G_{f_2}(x)$. Then, the ratio between the expected length of the overlapped frequency band and $\omega_1$ is: 
 	\begin{align}
 		\mathcal F(f_{2,\text{mn}},&f_{2,\text{mx}},\omega_2,\omega_1,f_1)\text{=}\int\nolimits_{0}^{\min\{\omega_1,\omega_2\}}\big[G_{f_2}( f_1\text{+}[\omega_1\text{+}\omega_2]/2\text{-}x)\nonumber\\
 		&-G_{f_2}(f_1{-}[\omega_1\text{+}\omega_2]/2\text{+}x)\big]/\omega_1 dx,\label{p1}
 	\end{align}
 \end{proposition}
 \begin{IEEEproof}
 	The CDF of length of the overlapped frequency band between packets from device 1 and 2 is:
 	\begin{align}
 		G_{v}(x)=&pr(v\le x)=pr(f_2\ge f_1+[\omega_1+\omega_2]/2-x)\nonumber\\
 		&\hspace{16mm}+pr(f_2\le f_1-[\omega_1+\omega_2]/2+x),\nonumber\\
 		=&1-G_{f_2}( f_1+[\omega_1+\omega_2]/2-x)\nonumber\\
 		&\hspace{2mm}+G_{f_2}(f_1-[\omega_1+\omega_2]/2+x).\label{p2}
 	\end{align}
 	Given $y$ as a random variable with CDF of $G_Y(y)$, mean of $y$ is derived as $\int [1-G_Y(y)]dy$ \cite{pap}. Using this fact, deriving \eqref{p1}  from \eqref{p2} is straightforward.
 \end{IEEEproof}
	When $f_2$ is uniformly distributed in $[f_{2,\text{mn}},f_{2,\text{mx}}]$, and 
 	$$f_{2,\text{mx}}>f_1+[\omega_1+\omega_2]/2\text{ and }  f_{2,\text{mn}}<f_1-[\omega_1+\omega_2]/2$$ hold, the ratio between  expected length of the overlapped frequency band and $\omega_1$ is
 	\begin{equation}\label{e1}
 		\mathcal F(f_{2,h},f_{2,\text{mx}},\omega_2,\omega_1,f_1)={ \omega_2}/({f_{2,\text{mx}}-f_{2,\text{mn}}}).\end{equation}
Finally, the average success probability over $f_j$ is derived as:
\begin{equation}\mathcal P_{\text{sc}}(j,d, \gamma_{\text{th}})=\int\nolimits_{f_{j,\text{mn}}}^{f_{j,\text{mx}}} \mathcal G_{f_j}(x)\mathcal P(j,d,\gamma_{\text{th}},x) dx,\label{hazf}
\end{equation}
in which $\mathcal G_{f_j}(x)=\partial G_{f_j}(x)/\partial x$ denotes the probability distribution function (PDF) of $f_j$ over $[f_{j,\text{mn}},f_{j,\text{mx}}]$.

 \subsection{Analytical Modeling of KPIs}\label{dr}
 First, we investigate the delay, i.e., the time span between packet generation at the device and successful    packet reception at the AP. Regarding the fact that each packet is transmitted with success probability $\mathcal P_{\text{sc}}(j,\gamma_{\text{th}})$, the average experienced delay for a successfully received packet from a class $j$ device is derived as:
 \begin{align}
 	D_j\text{=}\sum\nolimits_{n=1}^{N_{\text{tx}}}\big[n T_j &\text{+}[n\text{-}1]T_{w_j}\big] \mathcal P_{\text{sc}}(j,\gamma_{\text{th}})\big[1\text{-}\mathcal P_{\text{sc}}(j,\gamma_{\text{th}})\big]^{n\text{-}1}, \label{dn}
 \end{align}
 where $N_{\text{tx}}$ denotes the maximum number of transmissions for a packet,  and $T_{w_j}$ denotes the average waiting time between two retransmissions for class $j$ devices. 
 
 Now, we investigate the battery lifetime, i.e., the time span between deployment of a class $j$ device with battery capacity $E_j$ until when it has its battery drained. Regarding the fact that the reporting period for a class $j$ device is $T_j$, the expected battery lifetime is derived as:
\begin{equation}L_j= {E_j}\mathcal T_j/{\mathcal E_{j,b}},\end{equation}
 where $\mathcal E_{j,b}$ represents the average energy consumption per reporting period. Also, $\mathcal E_{j,b}$ can be modeled as:
 $${{\cal E}_{j,b}} =  {{P_c}{T_{{a_j}}}}  +  {[{P_c} \text{+} \eta {P_j}]{T_j}{{\bar n}_j}}  +  {[{{\bar n}_j} \text{-} 1][{P_c}{T_w} \text{+} {P_r}{T_{{\rm{ack}}}}]}  +  {{P_r}{T_{{\rm{ack}}}}},$$
 where the first term denotes energy consumption in data gathering/processing, the second term indicates the energy consumption in data transmission, the third term indicates energy consumption in listening for ACK and waiting for retransmission, and the fourth term indicates energy consumption in receiving acknowledgment. 
 Also, $P_c$ is a constant energy consumption in circuits, $\eta$ is the inverse power amplifier efficiency, $P_i$ is the transmit power, $T_{a_j}$ is the active time for data gathering/processing, $T_w$ is the waiting time for receiving ack, $P_r$ is the energy consumption in receiving data, $T_{\text{ack}}$ is the time-length of ack, and  $\bar n_j$ denotes the average number of transmissions for a successful packet transfer, as follows:  
 \begin{equation}\bar n_j=\sum\nolimits_{n=1}^{N_{\text{tx}}}n \mathcal P_{\text{sc}}(j,\gamma_{\text{th}})\mathcal P_{\text{ack}}\big[1-\mathcal P_{\text{sc}}(j,\gamma_{\text{th}})\mathcal P_{\text{ack}}\big]^{n-1},\label{bn}\end{equation}
 where $\mathcal P_{\text{ack}}$ represents the probability of successful ACK reception. Considering neighboring APs  as interfering nodes that their downlink transmissions can collide, $ \mathcal P_{\text{ack}}$ can be found from \eqref{hazf} for given density, transmit power, and communication characteristics of the APs. 
 Inserting $\mathcal P_{\text{sc}}$ from \eqref{hazf} into \eqref{bn}, we see  how the increase in the number of coexisting devices operating in grant-free mode, decreases the success probability, increases the average number of retransmissions, and hence decreases the battery lifetime.
 
%

 \subsection{Application to LoRa}

\subsubsection{LoRa Technology}
\begin{figure}[t!]
%
 	\centering
 	\includegraphics[scale=.65]{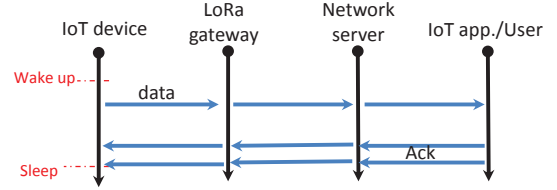}
 	\caption{Access protocol exchanges in LoRa}
 	\label{dlor}
 \end{figure}

 The LoRa wide area network provides seamless interoperability among IoT devices without any complex local installation requirements. LoRa is deployed in a star topology or cellular architecture in which a device  is connected to a central network server via  access points (APs) with the access protocol as depicted in \figurename \ref{dlor}. The LoRa network manages spreading factors (SFs) for each device in order  to optimize for the fastest possible data rate, which maximizes the network capacity. LoRa is utilizing chirp spread spectrum (CSS) as a modulation to maintain the immunity against the severe interference on unlicensed bandwidth and is fairly robust to multi-path fading and Doppler shifts \cite{LoRa2015Modulation}.  The high resilience to the interferers is key to operate efficiently in the public ISM band.  The main feature of CSS is that signals with different SFs can be distinguished and received simultaneously, even if they are transmitted at the same time on the same channel \cite{bankov2017mathematical}. \text{SF}s, ranging from 7 to 12, denote the number of chirps used to encode a bit. The higher chirp rate is, the better reconstruction of the received signal is attained, however, it stretches the  transmission time \cite{blenn2017lorawan}. For uplink transmission,  the duty-cycle is $ 1\% $ in EU 868 \cite{adelantado2016understanding}, and  devices use unslotted random access, similar to ALOHA. Downlink transmission can be done only during dedicated time intervals called receive windows, which follows successful uplink transmissions \cite{bankov2016limits}.

\subsubsection{Analysis of LoRa}
Assume  LoRa devices and APs have been deployed in a 2-dimensional space, i.e., no inter-technology interference is considered.  To get benefit from \eqref{hazf} for deriving the KPIs, we need to specify the time and frequency activity factors, i.e., $\xi_{i,j}$ and $\upsilon_{i,j}$ respectively. It is straightforward  that $\upsilon_{i,j}$ is 1, due to the intra-technology interference.  Denoting the number of available channels in LoRa with  $\mathcal C$, and the number of spreading factors (SF) available per channels as $|\text{SF}|$, the time activity factor is modeled as:
\begin{eqnarray}
\xi_{i,j} =  {1}/{\mathcal C}\times {1}/{|\text{SF}|}\times T_i/\mathcal T_i\label{prob},
\end{eqnarray}
where  $\mathcal T_i $ represents  the average time between generation of two successive packets at a device, and $T_i$ represents the average transmission time for each packet.
In LoRa protocol, devices wait for ACK within two windows after transmissions. If a node does not receive any ACK within the specified ACK windows, it transmits the packet again. ACKs in LoRa can be sent over a separate channel which is not used for transmission of devices' packets, and hence, the collision between data and ACK packets can be ignored. In this sense, the ACK success probability is as follows:
\begin{eqnarray}
\mathcal P_{\text{ack}} = \mathcal P_{\text{ack},1} + \mathcal P_{\text{ack},2}-\mathcal P_{\text{ack},1}\mathcal P_{\text{ack},2}
\end{eqnarray}
where $ \mathcal P_{\text{ack},1}$ and  $\mathcal P_{\text{ack},2}$ are probabilities of receiving ACK in the first and second receiving windows, and can be derived as described in Section \ref{dr}. 
Substituting \eqref{prob} in \eqref{hazf},  we can derive the success probability, i.e the probability of receiving   packet at the AP successfully. Subsequently, one can insert the  success probability in \eqref{dn} and \eqref{bn} in order to derive the delay expression. 
%

 \subsection{Joint Reception}\label{mul}
 Since in grant-free access, there are no established connections between devices and APs, multiple APs may receive the signal from a device. Therefore, the received signals at the different APs can be utilized to improve the received SINR and consequently eliminate the destructive effect of interfere technologies. To do so, each AP sends the received signal to the IoT server where the received signals are combined using combining methods such as MRC.  
  \subsubsection{Analysis of Joint Reception} 
  Denote distance between a class $j$ device and $m$th neighbor AP as $d_m$, where $m\in~\{1,\cdots,\mathcal M\}$.  
  Define event $e_m = \{ \text{SINR}\geq \gamma_{th} \}$ at AP $ m $, the event that the received SINR is greater than $\gamma_{th}$ at AP $m $. 
  We further extend the scenario to the case in which APs do not perform the decoding themselves but send the received signals to the network server for further processing and  joint reception. In this case,
 $$\mathcal P_{sc} (j,\mathop{\rm d},\gamma_{\text{th}},f_j)= 
 pr(H(\Gamma,\Pi)\ge \gamma_{\text{th}})=1-G_{H(\Gamma,\Pi)}(\gamma_{\text{th}}),$$
 in which $H(\Gamma,\Pi)$ is the function that describes the SINR gain achieved by combining, $G_{H(\Gamma,\Pi)}(x)=pr( H(\Gamma,\Pi)\le x)$ denotes CDF of $H(\Gamma,\Pi)$, $p^{\text{av}}_{i,j}$ denotes the fraction of time AP $i$ is in the listening mode, and 
 $$\Gamma= [\gamma_{1,j},\cdots,\gamma_{\mathcal M,j}]; \Pi=[p^{\text{av}}_{1,j},\text{...},p^{\text{av}}_{\mathcal M,j}].$$
 An upper-bound on $H(\cdot)$ is achieved by maximum ratio combining (MRC), i.e.,
 $$H(\Gamma,\Pi)=\sum\nolimits_{m=1}^{M}p^{\text{av}}_{m,j}\gamma_{m,j}.$$
 The CDF and PDF of $p^{\text{av}}_{m,j}\gamma_{m,j}$ are derived as \cite{pap}:
 $$G_{p^{\text{av}}_{m,j}\gamma_{m,j}}(x)=1-\mathcal P_{\text{sc}}(j,d_m,x/p^{\text{av}}_{m,j},f_j),$$
 $$ \mathcal G_{p^{\text{av}}_{m,j}\gamma_{m,j}}(x)= \partial G_{p^{\text{av}}_{m,j}\gamma_{m,j}}(x)/\partial x,$$
 in which $\mathcal P_{\text{sc}}$ has been found in \eqref{PsuccessCalc0_R}. Then, $G_{H(\Gamma,\Pi)} (x)$ is found as:
 $$G_{H(\Gamma,\Pi)} (x)= G_{p^{\text{av}}_{1,j}\gamma_{1,j}}*\mathcal G_{p^{\text{av}}_{2,j}\gamma_{2,j}}*\cdots* \mathcal G_{p^{\text{av}}_{\mathcal M,j}\gamma_{\mathcal M,j}}(x),$$
 in which $*$ denotes the convolution. Now, the success probability is upperbounded as follows:
 \begin{equation}
 \mathcal P_{\text{sc}}(j,d, \gamma_{\text{th}})=
 \int\nolimits_{f_{j,\text{mn}}}^{f_{j,\text{mx}}} \mathcal G_{f_j}(x) 
 [1-G_{H(\Gamma,\Pi)}(\gamma_{\text{th}})\big|_{ {f_j}=x} ]  dx,\label{eqo}
\end{equation}
which can be evaluated given statistics of the interference, i.e., $\mathcal G_{f_j}(x)$. Now, by substituting \eqref{eqo} in \eqref{dn} and \eqref{bn}, one can derive the delay and lifetime expressions.

 \section{Performance evaluations}\label{sectionIV}
\begin{figure}[t!]
	\centering
	\includegraphics[width=3.5in]{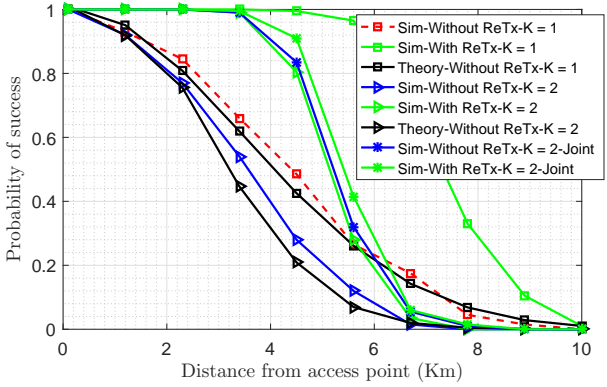}
	\caption{Impact of joint reception and coexistent technologies}
	\label{Psuccess}
\end{figure}

\begin{figure}[t!]
	\centering
	\includegraphics[width=3.5in]{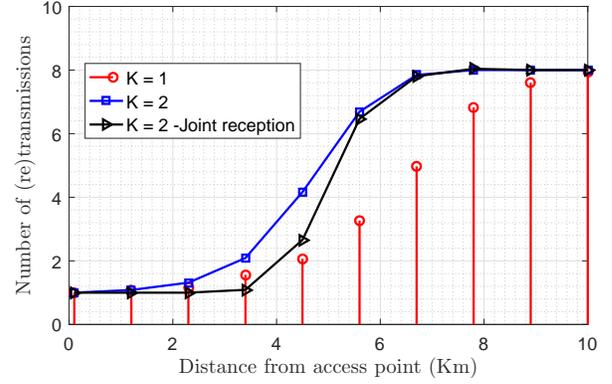}
	\caption{Impact of joint reception on the number of (re)transmissions}
	\label{NRetx}
\end{figure}

  In this section, we evaluate performance of a reference grant-free technology (RT), a simplified form of LoRa, while an interfering technology (IT) is active within the same frequency band. Assume we have  APs  and devices distributed over coverage area with PPP distribution. The transmit powers of IoT devices that belong to reference and competing technologies are set to $ 20 $ and $ 14 $  dBm.  We assume that the reference technology has 3 channels in which devices can choose to send their packets. ACKs are sent over a separate channel so we do not have any collision between packets and ACKs. The pathloss exponent, $\alpha $, is $ 4 $ and we have Rayleigh fading with unit mean exponential distribution. In case of collision, each packet can be retransmitted 7 times. A packet transmission  is successful if the SINR at the receiver is above the threshold, i.e., $ \gamma_{th} = 3$ dBm (after taking spreading gain into account). For joint reception, we assume  3 nearest access points can receive the signal and send it to a central unit in which, the received signals are combined using maximum ratio combining scheme. Other simulation parameters can be found in Table \ref{simR}.
   \begin{table}[t!]
   	\centering \caption{Simulation parameters.}\label{simR}
   	\begin{tabular}{p{4 cm}p{3 cm}}\\
   		\toprule[0.5mm]
   		{\it Parameter}&{\it Value}\\
   		\midrule[0.5mm]
   		Signal bandwidth &125 KHz\\
   		Noise power density &-174 dBm/Hz\\
   		Transmit power: RT, IT .& 20, 14 dBm\\
     	Frequency activity factor  & $ 10^{-1} $\\
     	Time activity factor& 0.01\\
     	Density of devices &   $ 10^{-2} $ devices/$ m^{2} $    \\
     Battery capacity& 4000 joule\\
        $ P_c, \gamma_{\text{th}}, \eta $& 100 mW, 3 dB, 0.7 \\
        $ T_a, T_{\text{ack}} $& 2, 1 sec\\
        $  \mathcal C, |\text{SF}|$& 3, 7 \\
   		\bottomrule[0.5mm]
   	\end{tabular}
   \end{table}
 In \figurename \ref{Psuccess}, the success probability as a function of distance to the serving  access point has been plotted. One sees that adding  coexisting technologies degrades the performance substantially. Also, we see how time diversity, i.e., retransmission, which is certainly achieved at the cost of a shorter device battery lifetime, increases the reliability of communications. We further observe that receiver diversity, 
 i.e., joint reception of  packets, which is achieved at the cost of more complex receivers, can significantly improve reliability of the system in presence of coexisting technologies reusing the same band. Also, one observes that there exist a point at which we cannot benefit from neither retransmissions (time diversity) nor joint reception (receiver diversity) to improve the performance, i.e., success probability. Thus, this is the point at which we need to change  the communication protocol to make it more robust to the interference.  In the case of LoRa, increasing the spreading factor brings more robust communications at the cost of longer transmission times, and hence, energy consumption per transaction. This observation pops out the idea to let each device to increase/decrease the spreading factor in use based on the received ACKs from the network.
The average required number of transmissions for having the packet  successfully received at the receiver has been demonstrated in \figurename \ref{NRetx}. One sees that having other active technologies on the same band  increases the number of  transmission retrials. One further observes that joint reception can well compensate the degradation caused by coexisting technologies, specially at points close to the access point where it outperforms the others due to the reinforced received signal strength. On the other hand, this reflects that by having both (i) denser networks, in which the average distance between a typical device and neighbor access points is not too large, and  (ii) using joint reception, one can significantly improve the network performance.

\begin{figure}[t!]
	\centering
	\includegraphics[width=3.5in]{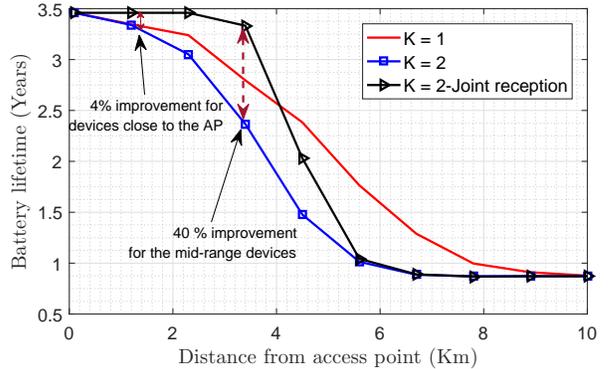}
	\caption{Battery lifetime performance}
	\label{Battery}
\end{figure}

In \figurename \ref{Battery}, the expected battery lifetime of a typical device as a function of distance to the access point has been depicted. As we expect, by increasing the distance, more retransmissions are needed, which degrade the battery lifetime. Also, similar to what observed in \figurename \ref{NRetx}, joint reception can significantly prolong the battery lifetime specially in short distances, i.e., in dense AP deployment scenarios. 

\begin{figure}[t!]
	\centering
	\includegraphics[width=3.5in]{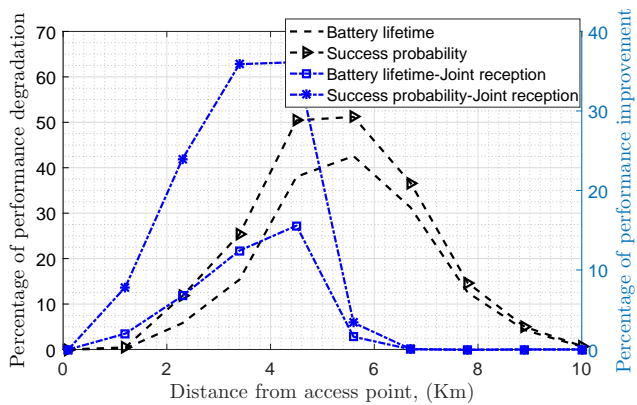}
	\caption{Percentage of performance degradation}
	\label{Performance}
\end{figure}

\figurename {\ref{Performance}} investigates the percentage of performance degradation in case of having multiple technologies active on the same band in comparison with the performance of single technology case. For devices close to the access point we do not see much performance degradation because the received signal strength is strong enough to decode the signal so the interference cannot affect the results heavily. For devices far away from the access point, the received signal is weak so that even in the absence of interferer the performance is very poor. The main impact of other technologies is observed for  devices that are neither too close nor too far from the access point. In this region, the battery lifetime and success probability (for single transmission) are reduced by about $ 40 $ \% and $ 50 $ \%, respectively. But by  joint reception, the destructive effect of  interferes can be removed.
In other words, reliability can be achieved by time diversity, i.e., increasing the number of retransmissions, as well as  by receiver diversity, i.e., by joint reception, where the former is achieved at the cost of shorter battery lifetimes for devices and the latter is achieved at the cost of increase in CAPEX and OPEX of the network, i.e., deploying more access points and more advanced receivers at the network server.  

 \section{Conclusion}\label{sectionV}
In this paper, we have presented an analytical model to evaluate the performance  of grant free IoT networks assuming that a radio spectrum is shared by competing radio access technologies. The model is developed with a cross layer approach which takes impacts of medium access control and physical layers into account.  We have derived closed-form expressions to compute the successful transmission probability, battery lifetime, and delay in the network. Analytical and simulation results  indicated  that existence of competing technologies could considerably degrade the performance. We have further demonstrated that using joint reception techniques, in which different access points receive the signal and relay it to the IoT server for joint reception, the performance degradation could be mitigated. Moreover,  simulation results confirmed that there exists  a distance from a typical access point, beyond where  neither the number of retransmissions nor the number of cooperating APs contribute to successful reception of data. Characterizing this region  is critical in system design, since to cover this region, one either needs to densify the network by deploying more APs, or increase the transmission power per data transfer at the device.    
  
  \ifCLASSOPTIONcaptionsoff
  \newpage
\fi

\bibliographystyle{IEEEtran}
\bibliography{biblios}
 \end{document}